\documentclass[aps,prl,twocolumn,showpacs,superscriptaddress,preprintnumbers]{revtex4-2}
\usepackage{amsmath}
\usepackage{epsfig}
\usepackage{color}
\usepackage{endnotes}
\usepackage{natbib}
\usepackage[colorlinks,breaklinks]{hyperref}
\usepackage{nicefrac}

\usepackage{bm}
\usepackage{dcolumn}
\usepackage{graphics}
\usepackage{amsmath}
\usepackage{amssymb}
\usepackage{color}
\usepackage{changes}
\newcommand{\bra}{\langle}
\newcommand{\ket}{\rangle}

\newcommand{\bs}[1]{\ensuremath{\boldsymbol{#1}}}

\newcommand{\be}{\begin{equation}}
\newcommand{\ee}{\end{equation}}
\newcommand{\bea}{\begin{align}}
\newcommand{\eea}{\end{align}}
\newcommand{\beqa}{\begin{eqnarray}}
\newcommand{\eeqa}{\end{eqnarray}}

\begin{document}

\preprint{LA-UR-23-27155}

\title{Short-range expansion for the quantum many-body problem}

\author{Ronen Weiss}
\thanks{ronenw@lanl.gov}
\affiliation{Theoretical Division, Los Alamos National Laboratory, Los Alamos, New Mexico 87545, USA}
\author{Diego Lonardoni}
\thanks{Present affiliation: XCP-2: Eulerian Codes Group, Los Alamos National Laboratory, Los Alamos, New Mexico 87545, USA}
\affiliation{Theoretical Division, Los Alamos National Laboratory, Los Alamos, New Mexico 87545, USA}
\author{Stefano Gandolfi}
\affiliation{Theoretical Division, Los Alamos National Laboratory, Los Alamos, New Mexico 87545, USA}

\date{\today}

\begin{abstract} 
In this work we derive a systematic short-range expansion of the many-body wave function. At leading order, the wave function is factorized to a zero-energy $s$-wave correlated pair and spectator particles, while terms that include energy derivatives and larger orbital angular momentum two-body functions appear at subleading orders. The validity of the expansion is tested for the two-body case, as well as the many-body case, where infinite neutron matter is considered. 
An accurate and consistent description of both coordinate-space two-body densities and the one-body momentum distribution is obtained.
These results show the possibility to utilize such an expansion for describing different observables in strongly-interacting many-body systems, including nuclear, atomic and condensed-matter systems. This work also enables a systematic description of large momentum transfer reactions in nuclear systems sensitive to short-range correlations, provides a link between such experiments and low-energy nuclear physics, and motivates measurement of new observables in these experiments. 
\end{abstract}

\maketitle

Non-relativistic quantum many-body systems are the focus of different research fields, including nuclear, atomic, and condensed-matter physics, and quantum chemistry. Studying such systems requires solving the many-body Schr\"odinger equation with reliable and systematic methods. Mean-field models are often used as a starting point for different perturbative approaches. However, in cases where strong correlations between the particles exist, these methods are usually ineffective. Significant contribution to such correlations usually arises due to strong interaction between particles at short distances.

In nuclear physics, for example, large short-range correlations (SRCs) prevent the use of many numerical methods. Methods based on renormalization-group (RG) techniques led to significant progress in the description of nuclei \cite{Heiko2020}, but they are mostly adequate to deal with relatively soft interactions. Quantum Monte Carlo methods are able to handle hard interactions, but are usually limited to relatively light systems \cite{Carlson:2014vla}. Developing a better theoretical understanding of SRCs is, therefore, important in order to make progress in the description of quantum many-body systems.

SRCs have been the focus of many experimental and theoretical studies in different fields, including atomic \cite{Tan08a,Tan08b,Tan08c,Braaten12,gandolfi11,Stewart10,sagi12,partridge05,werner09,Kuhnle2010,Bazak20} and nuclear systems \cite{Frankfurt88,Atti:2015eda,Hen:2016kwk,Arrington:2022sov}.
It was generally revealed that the interaction of two particles at short distances inside a many-body system results in correlated pairs with high relative momentum in a back-to-back configuration, that behave as an isolated two-body system.
Different properties of SRC pairs in nuclei, including their abundance and momentum dependence, were studied in detail. 
Nevertheless, there is still no systematic framework for describing SRCs and utilizing our experimental and theoretical knowledge of SRC properties for the description of more general observables that are affected by both long-range and short-range physics.
The purpose of this work is to derive a short-range expansion of the quantum many-body wave function that connects short-range and long-range physics together in an effort to develop such a framework. 

We start with the two-body case, considering a two-body eigenstate with energy $E$ and additional quantum numbers $\alpha$, denoted by $\varphi^E_\alpha(\bs{r})$.
$\varphi^E_\alpha(\bs{r})$ obeys the Schr\"odinger equation
\be
\left[ -\frac{\hbar^2}{2\mu}\nabla^2 + V(\bs{r})\right] \varphi^E_\alpha(\bs{r})  = E \varphi^E_\alpha(\bs{r}),
\ee
where $\mu$ is the reduced mass, $V$ is a two-body potential and $\bs{r}$ is the relative coordinate of the pair. At short distances, the kinetic energy term is dominant compared to $E$, and, therefore, the function does not depend on the value of $E$. We can thus conclude that
\be
\varphi^E_\alpha(\bs{r}) \xrightarrow[r \rightarrow 0]{} \varphi^{E=0}_\alpha(\bs{r}),
\ee
i.e., $\varphi^E_\alpha(\bs{r})$ coincides with the zero-energy eigenstate at short distances.
The zero-energy wave function can be identified as the leading term in a Taylor expansion around $E=0$ 
\be \label{eq:expan_2b}
\varphi^E_\alpha(\bs{r}) = \varphi^{E=0}_\alpha(\bs{r}) + \left(\frac{d}{dE} \varphi^{E=0}_\alpha(\bs{r})\right)E +
\frac{1}{2!} \left( \frac{d^2}{dE^2} \varphi^{E=0}_\alpha(\bs{r})\right)  E^2  + ... 
\ee
This Taylor expansion is a short-range expansion, because $\varphi^E_\alpha(\bs{r})$ does not depend on E at short distances and, therefore, terms involving energy derivatives vanish for $r \rightarrow 0$. As more terms are included, it is expected to describe $\varphi^E_\alpha(\bs{r})$ at larger and larger distances.

\begin{figure}\begin{center}
\includegraphics[width=\linewidth]{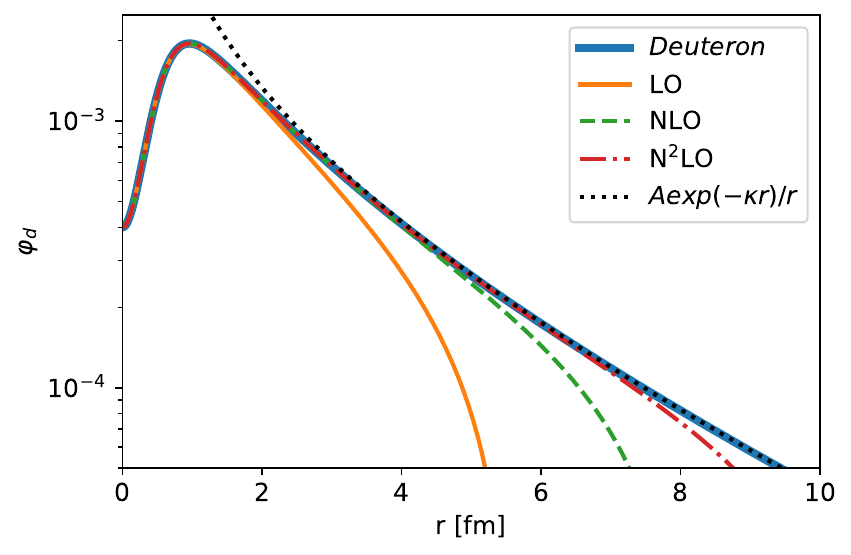}
\caption{\label{fig:E_expan_AV4P_deut}
The deuteron bound-state wave function using the AV4' potential compared to the short-range expansion of Eq. \eqref{eq:expan_2b}. 
The LO, NLO, and N$^2$LO terms of the expansion are shown. The dotted line shows the exponential long-range decay of the deuteron, where $\kappa = \sqrt{m|E|/\hbar^2}$.}
\end{center}\end{figure}

These claims can be tested against exact numerical calculations. We consider in Fig. \ref{fig:E_expan_AV4P_deut} the nuclear two-body bound deuteron. The AV4' potential \cite{Wiringa:2002} is used for simplicity, as it does not induce coupled channels.
It is a central potential in each of the 4 two-body spin-isospin channels. The leading order (LO), next-to-leading order (NLO) and next-to-next-to-leading order (N$^2$LO) expressions of the expansion of Eq. \eqref{eq:expan_2b} are shown in Fig. \ref{fig:E_expan_AV4P_deut} (where, e.g., terms with up to two energy derivatives are included at N$^2$LO).
As more terms in the expansion are included, the wave function approaches that of the detueron at larger and larger distances, reaching an excellent agreement up to $r\approx 7$ fm at N$^2$LO. 
We stress that there are no fitting parameters here, as we use the exact binding energy of the deuteron in Eq. \eqref{eq:expan_2b}. Fig. \ref{fig:E_expan_AV4P_deut}  also includes a line showing the long-range exponential decay of the deuteron, which agrees with the full wave function for $r \gtrsim 3$ fm. Combining the short-range expansion with the known long-range behavior, we obtain a complete description of the deuteron at all distances. 

We now move to the many-body case, considering an antisymmetric eigenstate of a given Hamiltonian with $A$ fermions $\Psi(\bs{r}_1,\bs{r}_2,...,\bs{r}_A)$ (we can deal with the bosonic case similarly). First, we expand $\Psi$ using the complete set of antisymmetric two-body eignestates $\left\{\varphi^E_\alpha(\bs{r}) \right\}$ of the same Hamiltonian
\be
\Psi(\bs{r}_1,\bs{r}_2,...,\bs{r}_A) = \sum_{\alpha,E} \varphi^E_\alpha(\bs{r}_{12}) A_\alpha^E(\bs{R}_{12},\bs{r}_3,...,\bs{r}_A).
\ee
Here, $\bs{r}_i$ is the single-nucleon coordinate of particle $i$, and $\bs{r}_{12}$ and $\bs{R}_{12}$ are the relative and center-of-mass (CM) coordinates of particles 1 and 2. The functions $A_\alpha^E$ serve as coefficients in this expansion. We note that this is an exact expansion and, therefore, $\Psi$ remains antisymmetric. 
Next, using the Taylor expansion of Eq. \eqref{eq:expan_2b}, we obtain
\begin{align} \label{eq:E_expan_Abody}
\Psi&(\bs{r}_1,\bs{r}_2,...,\bs{r}_A) = 
\sum_\alpha \varphi^{E=0}_\alpha(\bs{r}_{12}) A_\alpha^{(0)}(\bs{R}_{12},\bs{r}_3,...,\bs{r}_A) +
\nonumber \\ &
\sum_\alpha \left(\frac{d}{dE}\varphi^{E=0}_\alpha(\bs{r}_{12})\right) A_\alpha^{(1)}(\bs{R}_{12},\bs{r}_3,...,\bs{r}_A)
+ \nonumber \\ &
\sum_\alpha \left(\frac{d^2}{dE^2}\varphi^{E=0}_\alpha(\bs{r}_{12})\right) A_\alpha^{(2)}(\bs{R}_{12},\bs{r}_3,...,\bs{r}_A)
+... ,
\end{align}
where
\be
A_\alpha^{(n)}(\bs{R}_{12},\bs{r}_3,...,\bs{r}_A) \equiv 
\frac{1}{n!} \sum_E E^n A_\alpha^E(\bs{R}_{12},\bs{r}_3,...,\bs{r}_A).
\ee

This is our short-range expansion for the many-body case. Like the two-body case, we expect that as we go to larger values of $r_{12}$, more terms should be included in the expansion. However, the many-body case is more complicated because when organizing the terms in order of importance, we need to consider both the number of energy derivatives and the pair quantum numbers given by $\alpha$. 
In the limit $r_{12} \rightarrow 0$, channels with $s$-wave component are dominant, and terms with energy derivatives are suppressed.
Therefore, assuming a single $s$-wave channel, we obtain
\be \label{eq:GCF_LO_fact}
\Psi(\bs{r}_1,\bs{r}_2,...,\bs{r}_A)  \xrightarrow[r_{12} \rightarrow 0]{}
\varphi^{E=0}_s(\bs{r}_{12}) A_s^{(0)}(\bs{R}_{12},\bs{r}_3,...,\bs{r}_A),
\ee
where the subscript $s$ denotes the two-body $s$-wave channel.
This is a factorized form of the many-body wave function at short distances.

An identical short-range factorization ansatz is the basis of the Generalized Contact Formalism (GCF) \cite{Weiss14,Weiss:2015mba,Weiss:2016bxw}. The GCF, developed as an extension of Tan's theory for the zero-range model \cite{Tan08a,Tan08b,Tan08c,Braaten12}, is an effective model used to describe nuclear SRCs and their impact on different nuclear structure properties and reactions \cite{Weiss:2015mba,Weiss:2016obx,Cruz-Torres2020,Weiss:2018tbu,schmidt20,Pybus:2020itv,Duer:2018sxh,weiss2020inclusive,CLAS:2020rue,patsyuk2021unperturbed,Weiss:2021rig,Weiss:2016bxw,Weiss14,Weiss_EPJA16}, including two-body densities, momentum distributions, electron-scattering cross sections, and neutrinoless double beta decay matrix elements. The same approach is also useful for the description of other systems, like the case of Helium atoms \cite{Bazak20}. The subleading terms that appear in Eq. \eqref{eq:E_expan_Abody} provide corrections to this factorization ansatz.

Within the GCF, contact parameters, that measure the probability of finding SRC pairs in a nucleus, are defined as \cite{Weiss:2015mba}
\be
C_{\alpha}^{00} = \frac{A(A-1)}{2} \bra A_\alpha^{(0)} | A_\alpha^{(0)} \ket.
\ee
We can now generalize this definition to account for subleading terms
\be \label{eq:new_contacts}
C_{\alpha}^{mn} = \frac{A(A-1)}{2} S_{mn} \left[ \bra A_\alpha^{(m)} | A_\alpha^{(n)} \ket
+ \bra A_\alpha^{(n)} | A_\alpha^{(m)} \ket \right],
\ee
where $S_{mn}=1/2$ if $m=n$ and $S_{mn}=1$ if $m\neq n$.

This expansion and definition of contacts can be used to describe different quantities. We start with the two-body density $\rho_2(r)$, i.e., the probability of finding two particles at relative distance $r$ in a given system. Based on Eqs. \eqref{eq:E_expan_Abody} and \eqref{eq:new_contacts}, we can write a short-range expansion for this density
\be \label{eq:rho2_expan}
\rho_2(r) = \sum_\alpha \sum_{m\leq n} \phi_\alpha^{(m)*}(r) \phi_\alpha^{(n)}(r)C_\alpha^{mn},
\ee
where $\phi_\alpha^{(n)} \equiv \frac{d^n}{dE^n}\phi^{E=0}_\alpha$ and $\phi_\alpha^E(r)$ is the radial part of the two-body functions (see more details in the supplemental materials).
The two-body density of a many-body wave function is given here using only two-body functions and numerical coefficients $C_\alpha^{mn}$. The contacts $C_\alpha^{mn}$ depend on the many-body state and, therefore, are generally not simple to calculate directly. One of the important features of this expansion is that the same contact parameters appear in the description of different quantities, so they can be extracted from one quantity and used to predict another. We will demonstrate it here.

In our expansion, the LO term involves the $s$-wave channel without energy derivatives and the contact $C_s^{00}$. Going to larger distances, two possible terms might be involved in NLO corrections: (i) the same $s$-wave channel with one energy derivative involving the $C_s^{01}$ contact, and (ii) a $p$-wave channel without energy derivatives, involving a contact parameter $C_p^{00}$. To understand whether these two contributions enter at the same order, we can analyze the two-body Schr\"odinger equation. Considering a central potential, an $s$-wave solution behaves as $\phi_s \sim 1$ at short distances. The first energy derivative behaves as $\phi_s^{(1)} \sim r^2$, indeed suppressed compared to $\phi_s$. A $p$-wave solution behaves as $\phi_p \sim r$. Therefore, both $\phi_s^* \phi_s^{(1)}$ and $|\phi_p|^2$ behave as $r^2$ at short distances. Hence, $C_s^{01}$ and $C_p^{00}$ are expected to enter together at NLO. Similarly, $\phi_s^{(2)} \sim r^4$, $\phi_p^{(1)} \sim r^3$, and $\phi_d \sim r^2$ ($d$-wave solution), so $|\phi_s^{(1)}|^2$, $\phi_s^* \phi_s^{(2)}$,  $|\phi_d|^2$ and $\phi_p^* \phi_p^{(1)}$ all behave as $r^4$. Therefore, $C_s^{11}$, $C_s^{02}$, $C_d^{00}$ and $C_p^{01}$ are expected to contribute at N$^2$LO (see more details in the supplemental materials).

\begin{figure}\begin{center}
\includegraphics[width=\linewidth]{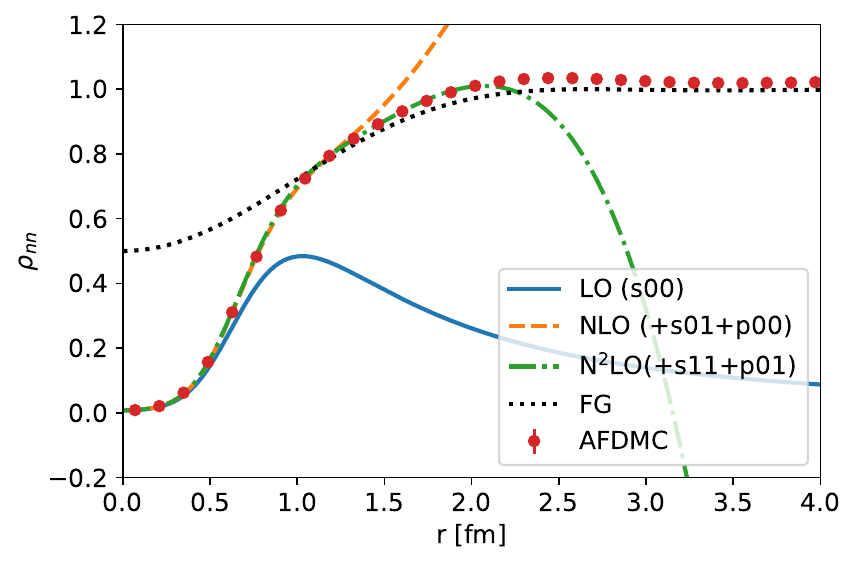}
\caption{\label{fig:rho_nn_PNM}
AFDMC calculations of $\rho_2(r)$ for infinite neutron matter at density 0.16 fm$^{-3}$ (red points), compared to the short-range expansion of Eq. \eqref{eq:rho2_expan}.  In the legend, the labels in parenthesis are of the form $\ell m n$, where $\ell$ is the orbital angular momentum quantum number. The FG result is also shown (black dotted line). See details in the text.}
\end{center}\end{figure}

With this understanding of the power counting, we can now test Eq. \eqref{eq:rho2_expan} against exact numerical calculations. For this purpose we will consider the two-body density of infinite neutron matter. 
We use the AV4' two-body interaction, together with the central UIX$_c$ three-body force \cite{Pudliner:1995wk}. Auxiliary-field diffusion Monte Carlo (AFDMC) calculations of $\rho_2(r)$ for infinite neutron matter at density of $0.16$ fm$^{-3}$ are compared to the short-range expansion in Fig. \ref{fig:rho_nn_PNM}.
At very short distances ($r \lesssim 0.5$ fm), the leading order $s$-wave contribution provides a good description. At larger distances, both $C_{s}^{01}$ and $C_p^{00}$ terms should be included, resulting in a good description for $r \lesssim 1.3$ fm. At N$^2$LO, $C_{s}^{11}$ and $C_p^{01}$ are included, extending the agreement to $r \lesssim 2$ fm.
In principle, at N$^2$LO we should also have contributions involving $C_s^{02}$ and $C_d^{00}$, but their $r$-dependence is very similar to $C_s^{11}$ for the AV4' potential, and, therefore, cannot be separated. 
The contact values are fitted to the AFDMC calculations (see more details in the supplementary).
At larger distances ($r \gtrsim 2$ fm), Fermi-gas (FG) calculations are in good agreement with the exact AFDMC calculations. Combining the short-range expansion at N$^2$LO level, which includes 5 terms and only two-body calculations, together with the long-range asymptotics based on the FG model, we obtain a good description of the many-body AFDMC calculations of neutron matter.

The above analysis of the two-body density shows the validity of our short-range expansion for the many-body case. The same expansion can also be used to describe other quantities. 
We can look at the one-body momentum distribution $n(k)$. Similarly to the derivation presented in Ref. \cite{Weiss:2015mba}, we obtain
\be \label{eq:mom_dist_expan}
n(k) = 2 \sum_\alpha \sum_{m\leq n} \tilde{\phi}_\alpha^{(m)*}(k) \tilde{\phi}_\alpha^{(n)}(k)C_\alpha^{mn},
\ee
where $\tilde{\phi}_\alpha^{(n)} \equiv \frac{d^n}{dE^n}\tilde{\phi}^{E=0}_\alpha$ and $\tilde{\phi}_\alpha^E(k)$ is the radial part of the two-body functions in momentum space. In this case, this expression should provide a high-momentum expansion of the one-body momentum distribution. We note that CM motion of the pair and three-body correlations are neglected here. They should become important around the Fermi momentum.

\begin{figure}\begin{center}
\includegraphics[width=\linewidth]{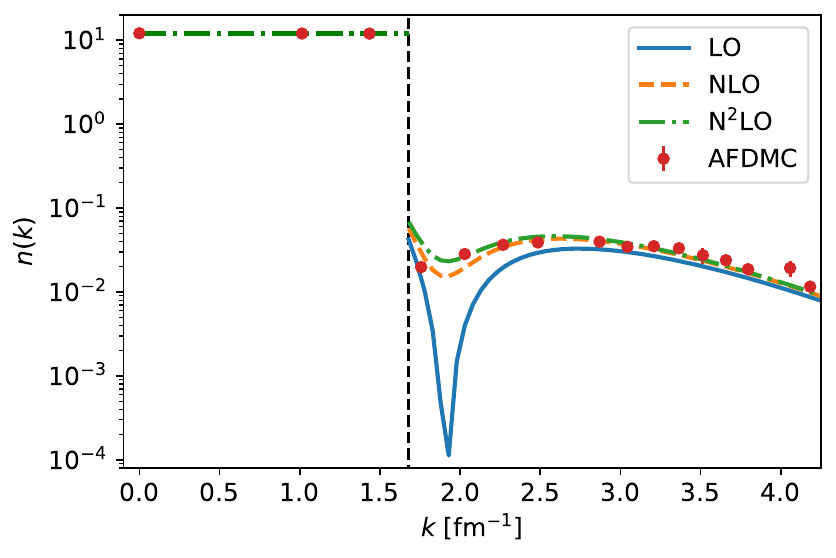}
\caption{\label{fig:n_k_PNM}
AFDMC calculations of $n(k)$ for infinite neutron matter at density 0.16 fm$^{-3}$ (red points), compared to the short-range expansion of Eq. \eqref{eq:mom_dist_expan} (above the Fermi momentum). The vertical dashed black line denotes the Fermi momentum. Below the Fermi momentum, a constant is fixed by normalization.}
\end{center}\end{figure}

Notice that the same contacts appear in both Eqs. \eqref{eq:rho2_expan} and \eqref{eq:mom_dist_expan}, as generally discussed above. Therefore, contact values fitted to the two-body density (Fig. \ref{fig:rho_nn_PNM}) can be used to test the expansion for the one-body momentum distribution and verify the consistency of the relations. This analysis is shown in Fig. \ref{fig:n_k_PNM}. 
We see that Eq. \eqref{eq:mom_dist_expan} provides a good description of the high-momentum part of the AFDMC one-body momentum distribution. The LO $s$-wave term describes the very high momentum tail, while next-order corrections are important at lower momenta, leading to a very good agreement with the exact calculations above the Fermi momentum. Specifically, the $p$-wave channel has significant contribution around $k=2$ fm$^{-1}$, where the $s$-wave function is zero. We can also describe the momentum distribution below the Fermi momentum as a constant fixed by the global normalization of $n(k)$, leading to a good description of the momentum distribution for all momentum values. This is relevant also in connection to a recent experimental study of the transition from mean-field to SRC domains~\cite{CLAS:2022odn}.

On top of the description of structure quantities, our expansion can be useful for analyzing reactions that are sensitive to SRCs, such as electron scattering in nuclear systems, neutron diffraction measurement of the static structure factor in liquid $^4$He \cite{Svensson:1980zz,Bazak20}, and radio-frequency spectroscopy in ultra-cold atomic systems \cite{Stewart10}.
For example, large momentum transfer electron scattering cross sections dominated by SRCs are well described by spectral function calculations which are based on the GCF LO short-range factorization of Eq. \eqref{eq:GCF_LO_fact} \cite{Weiss:2018tbu,schmidt20,Pybus:2020itv,Duer:2018sxh,CLAS:2020rue,patsyuk2021unperturbed}. No other methods are currently available to describe these reactions, beyond the very light nuclei. The above expansion provides next-order corrections for such spectral function calculations. They are important especially for relatively low momentum, similarly to the case of the momentum distribution (Fig. \ref{fig:n_k_PNM}). Therefore, it will provide a systematic description of such experiments and should allow us to extract contact values from experiments, including subleading terms. 
This will be the focus of future studies. We note that measurements of the spin and/or angular momentum of knocked-out pairs in such experiments will allow to separate the different contributions, e.g., the $s$-wave and $p$-wave contributions, and enable a more accurate extraction of contact values. Such contact values can then be used to predict different quantities for the same system.

\begin{figure} \begin{center}  
    \begin{tikzpicture}
        \node[anchor=south west,inner sep=0] (image) at (0,0) {\includegraphics[width=\linewidth]
{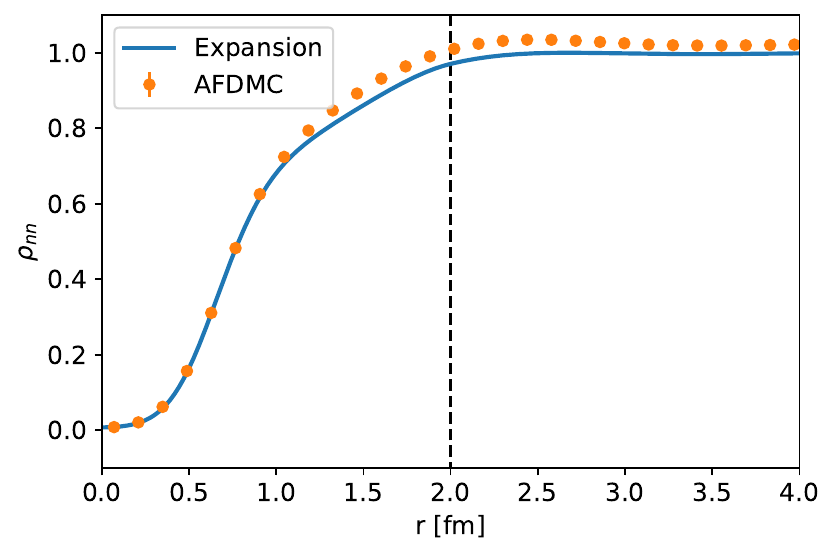}};
        \begin{scope}[x={(image.south east)},y={(image.north west)}]
            \node[anchor=south west,inner sep=0] (image) at (0.38,0.18) {\includegraphics[width=4.9cm]{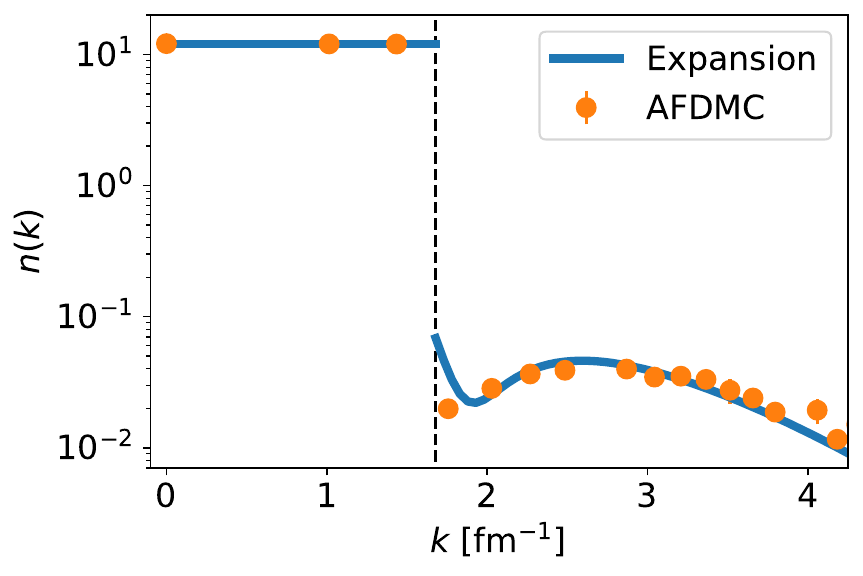}};
        \end{scope}
    \end{tikzpicture}
    \caption{\label{fig:rho_nn_matchFG} 
AFDMC calculations of $\rho_2(r)$ for infinite neutron matter at density 0.16 fm$^{-3}$ (orange points), and the expansion of Eq. \eqref{eq:rho2_expan}. The leading $s$-wave contact is fitted and the remaining 4 contact parameters are fixed by matching to the FG expression. The matching point is shown by the dashed line. For larger distances the FG density is used. Inset: AFDMC calculations of $n(k)$ for the same system and the expansion of Eq. \eqref{eq:mom_dist_expan} with the same contact values used in the main figure. Below the Fermi momentum, indicated by the dashed line, a constant is fixed by normalization.
}
\end{center} \end{figure}

The fact that our short-range expansion combined with long-range models (like the FG model) allows us to obtain a good description at all values of relative distances and momenta opens the possibility for describing different quantities that are affected by both mean-field physics and short- and long-range correlations. For example, we should be able to obtain a good description of the kinetic energy and two-body potential energy. 
Notice that at N$^2$LO level, our expansion requires 5 contact parameters for neutron matter, but this number can be reduced. Assuming that we know the value of the leading contact parameter ($C_s^{00}$), we can fix the values of the remaining 4 contact parameters by requiring continuous and smooth matching with long-range FG description of both spin-zero and spin-one two-body densities. This approach leads to a good description of both the two-body density and one-body momentum distribution, similar to the description shown in Figs. \ref{fig:rho_nn_PNM} and \ref{fig:n_k_PNM}, obtained by fitting all 5 contact parameters, see Fig. \ref{fig:rho_nn_matchFG}. Using this description of the densities, we obtain a two-body per-particle potential energy of $-29.3$ MeV (with the AV4' potential) and kinetic energy of $43.2$ MeV for the above neutron matter system. This is very close to the values obtained in AFDMC calculations, $-30.1$ MeV for the potential energy and $43.3$ MeV for the kinetic energy. We can see that with only knowledge of a single parameter (the leading order contact), the short-range expansion allows us to accurately calculate the kinetic energy and potential energy of the system.

To summarize, we have presented a systematic short-range expansion of quantum many-body wave functions. 
Description of observables is obtained using two-body functions and corresponding contact parameters. The same contacts are relevant for different quantities of a given system, allowing us to extract their values using one quantity and then predict others. We have identified an appropriate power counting by analyzing analytically the two-body problem. This expansion is relevant for various strongly-interacting many-body systems.

We have tested our approach against many-body ab-initio numerical calculations, considering nuclear systems as an example. Combined with asymptotic long-range models, a good description of both two-body density and one-body momentum distribution is obtained at all distances and momenta, enabling calculations of quantities that involve both long-range and short-range physics, like total potential energy and kinetic energy. We have also demonstrated the consistency of the different relations as a good description of the one-body momentum distribution with a clear order-by-order convergence is obtained using contact values fitted to the two-body densities. 

This work also provides a systematic framework for the analysis of large momentum transfer electron scattering experiments, focused on SRC physics, connecting them to low-energy nuclear physics studies. It provides next order corrections to the description of such experiments, including an important $p$-wave contribution, and motivates new experiments, such as measurements of the spin or orbital angular momentum of the outgoing pair, to isolate the contribution of different channels.

We would like to thank J. Carlson, J. Martin, S. Novario, R. Somasundaram, I. Tews, N. Barnea, B. Bazak, S. Beck, O. Hen, and E. Piasetzky for helpful discussions.
The work of R.W. was supported by the Laboratory Directed Research and Development program of Los Alamos National Laboratory under project number 20210763PRD1.
The work of S. G. was supported by U.S. Department of Energy, Office of Science,
Office of Nuclear Physics, under Contract No. DE-AC52-06NA25396,
by the DOE NUCLEI SciDAC Program, and by the DOE Early Career Research Program. 
Computer time was provided by the Los Alamos National Laboratory Institutional Computing Program, which is supported by the U.S. Department of Energy National Nuclear Security Administration under Contract No. 89233218CNA000001.

\bibliography{references}

\end{document}